\documentclass[11pt,a4paper]{article}
\usepackage{jcappub}

\begin{document}
\title {Observational Constraints on  Modified Chaplygin Gas from Cosmic Growth}

\author[a]{B. C. Paul,} 
\author[b]{P. Thakur}

\affiliation[a]{Department of Physics, North Bengal University \\
Siliguri, Dist: Darjeeling, Pin- 734013,  India}

\affiliation[b]{ Department of Physics, Alipurduar College \\
Alipurduar, Dist:Jalpaiguri,  Pin - 736122, India}

\emailAdd{bcpaul@iucaa.ernet.in} 
\emailAdd{prasenjit\textunderscore thakur1@yahoo.co.in}  

\abstract {We investigate the linear growth  function for   the large scale structures of the universe considering  modified Chaplygin gas  as  dark energy. Taking into account  observational growth  data  for a given range of redshift from the Wiggle-Z measurements  and rms mass fluctuations from Ly-$\alpha$ measurements  we numerically analyze cosmological  models to constrain the parameters of the MCG. The observational data of  Hubble parameter with redshift ($Z$) is also considered.   The Wang-Steinhardt ansatz for  growth index $\gamma$ and growth function $f$ (defined as $f=\Omega_{m}^{\gamma} (a)$) are considered for the numerical analysis.   The best-fit values of the equation of state parameters obtained here is employed to study  the  growth function $(f)$, growth index ($\gamma$) and state parameter ($\omega$) with redshift  $z$.  We note that MCG satisfactorily accommodates an accelerating phase followed by a matter dominated universe.}


\maketitle
\flushbottom

\section{Introduction}
\label{sec:introd}

Recent  cosmological observations  from supernova  \cite{mut,sb,sn1,sn2,sn4}, WMAP \cite{wmap,wmap2,hin,kog,sper}, BAO oscillation data \cite{bao}  predicted that   the present universe is passing through a phase of accelerating expansion which might have fuelled due to 
existence of a new source called  dark energy. In cosmological observations the expansion rate $H(z)$ is related to  various redshift which are  employed to obtain different parameters {\it e.g.}, distance modulus parameter is one of them. Although the analysis provide us with a satisfactory understanding of  cosmological dynamics it does not give a complete understanding of the evolution of the universe. Consequently additional observational inputs namely,  cosmic growth  of the inhomogeneous part of the universe for its structure formation are considered. The growth of the large scale structures derived from linear matter density contrast $\delta(z)\equiv\frac{\delta\rho_{m}}{\rho_{m}}$ of the universe is considered an  important tool in constraining cosmological model parameters. To describe the evolution of the inhomogeneous energy density it is preferable to parametrize the growth function $f=\frac{d\log\delta}{d\log a}$ in terms of growth index ($\gamma $). Initially Peebles \cite{peebles} and subsequently Wang and Steinhardt \cite{wangstein} parametrized $\delta$ in terms of $\gamma$ to obtain cosmologies which are useful in  different contexts discussed in the literature \cite{linder,las,ja,hu,lue,ac,lue3, koivisto,dan,so}. It is possible to understand  dark energy in cosmology   from the analysis of  observational data from the observed  expansion  rate ($H(z)$) and growth of matter density contrast $\delta(z)$ data simultaneously.\\

It is known that in general theory of relativity ordinary matter fields available from standard model of particle physics fails to account the present observations. As a result  modification of  the matter sector of the  Einstein-Hilbert action with exotic matter is considered in the literature.
  Chaplygin gas (CG) is considered to be one such candidate for dark energy.  The equation of state (henceforth, EoS) for CG  is 
\begin{equation}
p=-\frac{A}{\rho}
\end{equation}
where $A$ is positive constant.
It may be important to mention here that the initial idea of CG originated in Aerodynamics \cite{chap}. But it is ruled out  in cosmology from observations. Subsequently  the equation of state for CG is generalized to incorporate   different aspects of the observational  universe.  The equation of state for generalized Chaplygin gas (in short, GCG) \cite{billic, bento} is given by
\begin{equation}
p=-\frac{A}{\rho^\alpha}
\end{equation}
with $ 0 \leq \alpha \leq 1 $. In the above EoS, Chaplygin gas corresponds to $\alpha=1$ \cite{chap}. It has two free parameters $A$ and $\alpha$. Initially GCG behaves like a pressureless fluid at the early stages of the evolution of the universe, but at a later stage it behaves like a  cosmological constant. In the context of string theory  Chaplygin gas emerges from the dynamics of a generalized d-brane in a (d+1,1) space time. It can be described by a complex scalar field which is obtained from a generalized Born-Infeld action. 
More recently modified Chaplygin gas (in short, MCG)  is considered in describing  dark energy because of its negative pressure \cite{debnath,liu,thakur}.  The equation of state for the MCG is given by:
\begin{equation}
\label{mcgeos}
p=B\rho-\frac{A}{\rho^\alpha}
\end{equation}
where $A$, $B$, $\alpha$ are positive constants
with $0 \leq \alpha \leq 1 $ . The above EoS reduces to that of GCG model \cite{billic,bento} when one sets  $B=0$. A cosmological constant $\Lambda$  emerged by setting  $\alpha = -1$ and $A = 1 + B$. If one considers $A = 0$, it reduces to an EoS which describes a perfect fluid with  $\omega = B$, {\it e.g.},  a quintessence model \cite{lxu}. MCG contains one more free parameter namely, $B$ over GCG..It may be pointed out here that MCG is a single fluid model which unify dark matter and dark energy. MCG behaves as dark matter when its energy density evolves as $\rho \propto a^{-3}$ (where $a(t)$ represents the scale factor) in the early epoch whereas the constant energy density behaves as dark energy  accommodating late time acceleration \cite{smat}.  The MCG is consistent with (i) Gravitational lensing test  \cite{silva, dev} and (ii) Gamma-ray bursts \cite{bertolami}.\\

In this paper we determine  constraints on EoS parameters of the MCG using different observational data namely, the growth function and growth index in a FRW universe.  The growth data and the Stern data set \cite{stern} of $H(z)$ vs. $z$
 are considered here for the analysis. The growth data given in Table -\ref{tab1}  consists of a number of data points within redshift range (0.15 to 3.8) which are  related to growth function $f$. It may be pointed out here that redshift that estimates the linear growth rate are considered  from various projects/surveys including the latest Wiggle-Z measurements. Gupta {\it et al.} \cite{anjan} obtained constrains on  GCG parameters using the above data. Cosmological model dominated by viscous dark fluid is also considered  in Ref.\cite{velt} where it is found that viscous fluid mimics as $\Lambda$CDM model when coefficient of viscosity varies as $\rho^{-1/2}$ providing excellent agreement both with supernova and $H-z$ data. The viscous cosmological model is found  analogous to GCG model. In addition to the above data other observational set of growth data given in (Table- \ref{tab2} from  various sources such as: the redshift distortion of galaxy power spectra \cite{hawkins1}, root mean square $(rms)$  mass fluctuation ($\sigma_{8}(z)$) obtained from galaxy and Ly-$\alpha$ surveys at various redshifts \cite{viel1,viel2}, weak lensing statistics \cite{kaiser}, baryon acoustic oscillations \cite{bao}, X-ray luminous galaxy clusters \cite{manz}, Integrated Sachs-Wolfs (ISW) Effect \cite{rees,am,kaiser3,cr,pog} etc. are important. It is known that redshift distortions are caused by velocity flow induced by gravitational potential gradient which evolved  due to the growth of the universe under gravitational attraction  and dilution of the potentials due to the cosmic expansion. The gravitational growth index $\gamma$ has considerable impact on the redshift distortion \cite{linder}. The  cluster abundance evolution,  however, found strongly depends on rms mass fluctuations ($\sigma_{8}(z)$) \cite{wangstein}.\\

We  adopt here chi-square minimization technique to constrain different parameters of the EoS for a viable cosmological model considering MCG. In the analysis total chi-square is constituted using the growth data and  the $(H(z)-z)$ OHD data. The best-fit values of the model parameters are then determined from the chi-square function to study evolution of the universe.
The paper is organized as follows : 
In sec.2,  we set up Einstein field equations. In sec.3,  growth index parametrization in terms of EoS parameters is given. In sec.4,  constraint on the EoS parameters obtained from observational data are presented.  Finally, we conclude in sec.5.

\section{Einstein Field Equations}

The Einstein field equation is given by
\begin{equation}
\label{ricci}
R_{\mu \nu}-\frac{1}{2} g_{\mu \nu} R = 8 \pi G \; T_{\mu \nu}
\end{equation}
where $R_{\mu \nu}$ represents Ricci tensor, $R$ represents Ricci scalar, $T_{\mu \nu}$ represents energy momentum tensor and $g_{\mu \nu}$ represents the metric tensor in 4-dimensions.
We consider a Robertson-Walker  metric which is given by
\begin{equation}
\label{metric}
ds^{2} = - dt^{2} + a^{2}(t) \left[ \frac{dr^{2}}{1- k r^2} + r^2 ( d\theta^{2} + sin^{2} \theta \;
d  \phi^{2} ) \right]
\end{equation}
where  $k=0,+1(-1)$ is the curvature parameter in the spatial section representing flat, closed (open) universe and $a(t)$ is the scale factor of the universe with $r,\theta,\phi$   co-moving co-ordinates.

Using metric (\ref{metric}) in the Einstein field eq. (\ref{ricci}),  we obtain the following equations:
\begin{equation}
\label{fried}
3 \left( \frac{\dot{a}^2}{a^2} + \frac{k}{a^2} \right) = 8 \pi G \; \rho, 
\end{equation}
\begin{equation}
2 \frac{\ddot{a}}{a} + \frac{\dot{a}^2}{a^2} + \frac{k}{a^2} = - 8 \pi G \; p,
\end{equation}
where $\rho$ and $p$ represent the energy density and pressure respectively. The conservation equation is given by
\begin{equation}
\label{energy}
\frac{d\rho}{dt} + 3 H \left(\rho + p \right) = 0, 
\end{equation}
where $H = \frac{\dot{a}}{a}$ is Hubble parameter.

Using EoS given by eq.(\ref{mcgeos}) in eq.(\ref{energy}),and integrating once one obtains  energy density for a modified Chaplygin gas which is given by 
\begin{equation}
\label{rhomcg}
\rho_{mcg}=\rho_{0}\left[A_{S}+\frac{1-A_{S}}{a^{3(1+B)(1+\alpha)}}\right]^{\frac{1}{1+
\alpha}}
\end{equation}
 where $ A_{S} = \frac{A}{1+B}\frac{1}{\rho_{0}^{\alpha +1}}$ with $ B\neq -1$, $\rho_o$ is an integration constant. The scale factor of the universe is related to the   redshift parameter ($z$) as $\frac{a}{a_{0}}=\frac{1}{1+z}$, where   we choose the present scale factor of the universe   $a_{0}=1$ for convenience. The MCG model parameters are  
$ A_{S}$, $ B$  and $\alpha$.
 From eq. (\ref{rhomcg}), it is evident that the positivity condition of the energy density is ensured when $0\leq A_{S} \leq 1$. From  eq. (\ref{rhomcg}), one recovers  the standard $\Lambda CDM$ model for  $\alpha= 0 $ and $B = 0$.
The Hubble parameter can be expressed as  a function of  redshift  using the field  eq. (\ref{fried}), which is given by 
\[
H(z)=H_0[\Omega_{b0}(1+z)^3
+ 
\] 
\begin{eqnarray}
\label{hpara}
\; \; \; \; \; (1-\Omega_{b0})
[(A_s+(1-A_s)(1+z)^{3(1+B)(1+\alpha)})^{\frac{1}{1+\alpha}}]]^{\frac{1}{2}}.
\end{eqnarray}
where $\Omega_{b0}$, $H_{0}$ represent the present baryon density and present Hubble parameter respectively.

The sound speed is given by
\begin{equation}
\label{sound}
c^2_{s}=\frac{\delta p}{\delta\rho}=\frac{\dot{p}}{\dot{\rho}}
\end{equation}
which reduces to
\begin{equation}
\label{sound1}
c^2_{s}=B+\frac{A_{s}\alpha(1+B)}{\left[A_{s}+(1-A_{s})(1+z)^{3(1+B)(1+\alpha)})\right]}.
\end{equation}
In terms of state parameter it becomes
\begin{equation}
\label{sound2}
c^2_{s}=-\alpha\omega+B(1+\alpha).
\end{equation}
It may be mentioned here that for causality and stablity under perturbation it is required to satisfy  the inequality condition $c_{s}^2 \leq 1$ \cite{lxu}.

\section{ Parametrization of the Growth Index}

 The growth rate of the large scale structures is derived from matter density perturbation $\delta=\frac{\delta\rho_{m}}{\rho_{m}}$ (where  $\delta\rho_{m}$ represents the fluctuation of matter density $\rho_{m}$) in the linear regime which satisfies
\begin{equation}
\label{grrate}
\ddot{\delta}+2\frac{\dot{a}}{a}\dot{\delta}-4\pi G_{eff}\rho_{m}\delta=0.
\end{equation}
The field equation for the background cosmology comprising both  matter and MCG in FRW universe are given below
\begin{equation}
\left(\frac{\dot{a}}{a}\right)^{2}=\frac{8\pi G}{3}(\rho_{b}+\rho_{mcg}),
\end{equation}
\begin{equation}
2\frac{\ddot{a}}{a}+\left(\frac{\dot{a}}{a}\right)^{2}=-8\pi G\omega_{mcg}\rho_{mcg}
\end{equation}
where $\rho_b$ represents the background energy density and $\omega_{mcg}$  represents the state parameter for MCG which is given by
\begin{equation}
\omega_{mcg}=B-\frac{A_{s}(1+B)}{\left[A_{s}+(1-A_{s})(1+z)^{3(1+B)(1+\alpha)})\right]}.
\end{equation}
 We now replace the time ($t$) variable to $ ln \, a$ in  eq.(\ref{grrate}) and obtain
\begin{equation}
\label{delta}
(\ln\delta)^{''}+(\ln\delta)^{'2}+(\ln\delta)^{'}
\left[\frac{1}{2}-\frac{3}{2}\omega_{mcg}(1-\Omega_{m}(a))\right]=\frac{3}{2}\Omega_{m}(a)
\end{equation}
where
\begin{equation}
\label{density}
\Omega_{m}(a)=\frac{\rho_{m}}{\rho_{m}+\rho_{mcg}}.
\end{equation}
The  effective matter density is given by  $\Omega_{m}=\Omega_{b}+(1-\Omega_{b})(1-A_{s})^{(1/1+\alpha)}$ \cite{li}.
Using the energy conservation eq. (\ref{energy}) and changing the variable from $\ln a$ to $\Omega_{m}(a)$ once again,  the eq. (\ref{delta}) can be expressed in terms of the logarithmic growth factor $f=\frac{d\log\delta}{d\log a}$ which is given by 
\begin{equation}
\label{gfactor}
3\omega_{mcg}\Omega_{m}(1-\Omega_{m})\frac{d f}{d \Omega_{m}}+f^{2}
 +f\left[\frac{1}{2}-\frac{3}{2}\omega_{mcg}(1-\Omega_{m}(a))\right]=\frac{3}{2}\Omega_{m}(a).
\end{equation}

The logarithmic growth factor $f$, according to Wang and Steinhardt   \cite {wangstein} is given by
\begin{equation}
\label{ansatzf}
f=\Omega_{m}^{\gamma}(a)
\end{equation} 
where $\gamma$ is the growth index parameter. In the case of flat dark energy model  with constant state parameter  $\omega_{0}$, the growth index $\gamma$ is given by
\begin{equation}
\label{gamflat}
\gamma=\frac{3(\omega_{0}-1)}{6\omega_{0}-5}.
\end{equation}
For a  $\Lambda CDM$ model, it reduces to $\frac{6}{11}$ \cite{linder,evl}, for a matter dominated model, one gets $\gamma=\frac{4}{7}$ \cite{fry,ne}. One can also write $\gamma$  as a parametrized function of redshift parameter $z$.  One such parametrisation is
$\gamma(z)=\gamma(0)+\gamma^{'}z$, with $\gamma^{'}\equiv\frac{d\gamma}{dz}|_{(z=0})$ \cite{polarski, gan}. It has been shown recently \cite{ishak} that the parametrization smoothly interpolates a low and intermediate redshift range to a high redshift range up to the cosmic microwave background   scale. The above parametrization is also taken up in different contexts \cite{dosset}.
In this paper we parametrize  $\gamma$  in terms of MCG parameters namely,  $A_{s}$, $\alpha$ and $B$.
 Therefore, we begin with the following  ansatz which is given by
\begin{equation}
\label{ansatzmcg}
f=\Omega_{m}^{\gamma(\Omega_{m})} (a)
\end{equation} 
where the growth index parameter $\gamma(\Omega_{m})$ can be expanded in Taylor series around $\Omega_{m} = 1$ as
\[
f=\gamma(\Omega_{m}) \hspace{8cm.}
\]
\begin{equation}
\label{tylor}
\;\; \; \; \; \; \; \;  =\gamma(\Omega_{m}-1)+(\Omega_{m}-1)\frac{d\gamma}{d\Omega_{m}}(\Omega_{m}=1) +O(\Omega_{m}-1)^{2}.
\end{equation}
Consequently eq.(\ref{tylor}) can be rewritten in terms of $\gamma$ as
\[
3\omega_{mcg}\Omega_{m}(1-\Omega_{m})\ln\Omega_{m}\frac{d\gamma}{d \Omega_{m}}-3\omega_{mcg}\Omega_{m}(\gamma-\frac{1}{2})+
\]
\begin{equation}
\label{tylomega}
\; \; \; \; \; \Omega_{m}^{\gamma}-\frac{3}{2}\Omega_{m}^{1-\gamma}+3\omega_{mcg}\gamma-\frac{3}{2}\omega_{mcg}+\frac{1}{2}=0.
\end{equation}
Differentiating once again the above equation around $\Omega_{m}=1$, one obtains  zeroth order term in the expansion  for $\gamma$ which is given by
\begin{equation}
\label{grmcg}
\gamma=\frac{3(1-\omega_{mcg})}{5-6\omega_{mcg}},
\end{equation}
it  agrees with dark energy model for a constant $\omega_{0}$ (eq. \ref{gamflat}). 
In the same way differentiating it twice and thereafter by a  Taylor expansion  around $\Omega_{m}=1$, one obtains the first order terms  in the expansion which is given by
\begin{equation}
\label{forder}
\frac{d\gamma}{d\Omega_{m}}|_{\Omega_{m}=1} = \frac{3(1-\omega_{mcg})(1-\frac{3\omega_{mcg}}{2})}{125(1-\frac{6\omega_{mcg}}{5})^{3}}.
\end{equation}
Substituting it in eq. (\ref{tylor}), $\gamma$ is obtained,  the first order term in this case is  approximated to
\begin{equation}
\label{gammamcg}
\gamma=\frac{3(1-\omega_{mcg})}{5-6\omega_{mcg}}+(1-\Omega_{m})\frac{3(1-\omega_{mcg})(1-\frac{3\omega_{mcg}}{2})}{125(1-\frac{6\omega_{mcg}}{5})^{3}}.
\end{equation}
Using the expression of $\omega_{mcg}$ in the above,  $\gamma$ can be parametrised with  MCG  parameters.
 We define normalised growth function $g$ from the numerically obtained solution using eq. (\ref{delta}) which is given by 
 \begin{equation}
 \label{g}
 g(z)\equiv\frac{\delta(z)}{\delta(0)}.
 \end{equation}
 The corresponding  normalised growth function obtained from the parametrized form of $f$  follows from eq.(\ref{ansatzmcg}) which  is given by
 \begin{equation}
 g_{th}(z)=\exp\oint\Omega_{m}(a)^{\gamma}\frac{da}{a}.
 \end{equation}
  The above expression will be  employed to construct  chi-square function in the next section.

\section{Observational Analysis}

\begin{table}[tbp]
\centering
		\begin{tabular}{|lr|c|c|}
		\hline
		z     & $f_{obs}$ &    $\sigma$ &    $Ref.$\\
		\hline
		$0.15$   & 0.51 &  0.11 &  \cite{hawkins,verde}\\
		$0.22$ & 0.60 & 0.10 &  \cite{blake}\\
		$0.32$ & 0.654 & 0.18 & \cite{reyes}\\
		$0.35$ & 0.70 & 0.18 &  \cite{tegmark}\\
		$0.41$ & 0.70 & 0.07 & \cite{blake}\\
		$0.55$ & 0.75 & 0.18 & \cite{ross}\\
		$0.60$ & 0.73 & 0.07 & \cite{blake}\\
		$0.77$ & 0.91 & 0.36 & \cite{guzzo}\\
		$0.78$ & 0.70 & 0.08 & \cite{blake}\\
		$1.4$ & 0.90 & 0.24 & \cite{angela}\\
		$3.0$ & 1.46 & 0.29 & \cite{mcdon}\\
		\hline	
		\end{tabular}
\caption{\label{tab1} Data for the observed growth functions $f_{obs}$ used in our analysis }	
\end{table}

\begin{table}[tbp]
\centering
		\begin{tabular}{|lr|c|c|}
		\hline
		z     & $\sigma_{8}$ &    $\sigma_{\sigma{_{8}}}$ &    $Ref$\\
		\hline
		$2.125$   & 0.95 &  0.17 &  \cite{viel1}\\
		$2.72$ & 0.92& 0.17 &       \\
		$2.2$ & 0.92 & 0.16 & \cite{viel2}\\
		$2.4$ & 0.89 & 0.11 &   \\
		$2.6$ & 0.98 & 0.13 &   \\
		$2.8$ & 1.02 & 0.09 & \\
		$3.0$ & 0.94 & 0.08 & \\
		$3.2$ & 0.88 & 0.09 & \\
		$3.4$ & 0.87 & 0.12 & \\
		$3.6$ & 0.95 & 0.16 & \\
		$3.8$ & 0.90 & 0.17 & \\
		$0.35$ & 0.55 & 0.10 & \cite{marin}\\
		$0.6$ & 0.62 & 0.12 &    \\
		$0.8$ & 0.71 & 0.11 &    \\
		$1.0$ & 0.69 & 0.14 &    \\
		$1.2$ & 0.75 & 0.14 &    \\
		$1.65$ & 0.92 & 0.20 &   \\
		\hline	
		\end{tabular}
\caption{\label{tab2} Data for the rms mass fluctuations ($\sigma_{8}$) at various redshift }	
\end{table}

The redshift distortion parameter $\beta$, is related to the growth function $f$  as $\beta=\frac{f}{b}$, where  $b$ represents the bias  factor connecting total matter perturbation ($\delta$)  and galaxy perturbations ( $\delta_{g}$ ) ($b=\frac{\delta_{g}}{\delta}$) \cite{blake,tegmark,ross,angela}. The data for $\beta$ and $b$ at various redshifts are taken from  Ref. \cite{blake,porto}.  In the present case we   analyze  cosmological models in the presence of  modified Chaplygin gas in respect of cosmic growth function.
As it is not possible to determine $\beta$  as is pointed out in  Ref. \cite{mcdon}, we use other power spectrum amplitudes of Ly-$\alpha$ forest data.
\begin{table}
  \centering
  \begin{tabular}{|lr|c|c|}
  \hline
  {\it z Data} & $H(z)$ & $\sigma$ \\
  \hline
   0.00 & 73  & $ \pm $ 8.0	 \\
   0.10 & 69  & $ \pm $ 12.0 \\
   0.17 & 83  & $ \pm $ 8.0 \\
   0.27 & 77  & $ \pm $ 14.0 \\
   0.40 & 95  & $ \pm $ 17.4 \\
   0.48 & 90  & $ \pm $ 60.0 \\
   0.88 & 97  & $ \pm $ 40.4 \\
   0.90 & 117 & $ \pm $ 23.0 \\
   1.30 & 168 & $ \pm $ 17.4 \\
   1.43 & 177 & $ \pm $ 18.2 \\
   1.53 & 140 & $ \pm $ 14.0 \\
   1.75 & 202 & $ \pm $ 40.4 \\

\hline
\end{tabular}
\caption{\label{tab3} $H(z) vs. z$ data from Stern {\it et al. } \cite{stern}}
\end{table} 

We define chi-square of the growth function $f$ as
\begin{equation}
\chi^{2}_{f}(A_{s},B,\alpha)=\Sigma\left[\frac{f_{obs}(z_{i})-f_{th}(z_{i},\gamma)}{\sigma_{f_{obs}}}\right]^{2}
\end{equation}
where $f_{obs}$ and $\sigma_{f_{obs}}$ are obtained from Table-\ref{tab1}. However,  $f_{th}(z_{i},\gamma)$ is obtained from eqs. (\ref{ansatzmcg}) and (\ref{gammamcg}).
 Another  observational probe for the matter density perturbation $\delta(z)$ is derived from the redshift dependence of the rms mass fluctuation $\sigma_{8}(z)$. The $rms$ mass fluctuation $\sigma_{8}(z)$ is defined as
\begin{equation}
\sigma^{2}(R,z)=\int_{0}^{\inf} W^{2}(kR)\Delta^{2}(k,z)dk/k
\end{equation}
where 
\begin{equation}
W(kR)=3\left(\frac{\sin(kR)}{(kR)^{3}}-\frac{\cos(kR)}{(kR)^{2}}\right)r,
\end{equation}
\begin{equation}
\Delta^{2}(kz)=4\pi k^{3}P_\delta(k,z),
\end{equation}
with $R=8h^{-1}$ Mpc. In the above  $P_\delta(k,z)\equiv(\delta^{2}_{k})$ represents the mass power spectrum at redshift ($z$). The function $\sigma_{8}(z)$ is  related to  $\delta(z)$ as
\begin{equation}
\sigma_{8}(z)=\frac{\delta(z)}{\delta(0)}\sigma_{8}|_{(z=0)}
\end{equation}
which implies
\begin{equation}
s_{th}(z_{1}, z_{2}) \equiv \frac{\sigma_{8}(z_{1})}{\sigma_{8}(z_{2})}=\frac{\delta(z_{1})}{\delta(z_{2})}=
\frac{\exp\left[ \int_{1}^{\frac{1}{1+z_{1}}}  \Omega_{m}(a)^{\gamma}\frac{da}{a} \right] }{\exp \left[ \int_{1}^{\frac{1}{1+z_{2}}}  \Omega_{m}(a)^{\gamma}
\frac{da}{a} \right]}.
\end{equation}
Currently available data points $\sigma_{8}(z_{i})$ are taken from the observed redshift evolution of the flux power spectrum of Ly-$\alpha$ forest \cite{viel1,viel2,marin}. Using these we define  a new chi-square  function which is given by
\begin{equation}
\chi^{2}_{s}(A_{s},B,\alpha)=\Sigma\left[\frac{s_{obs}(z_{i},z_{i+1})-s_{th}(z_{i},z_{i+1})}{\sigma_{s_{obs,i}}}\right]^{2}.
\end{equation}
 For the numerical analysis we use  data given in   Table- \ref{tab2} .
From the Hubble parameter vs. redshift data \cite{stern} we define another chi-square $\chi^2_{H-z}$ function which is given by
\begin{equation}
\chi^{2}_{H-z}(H_{0},A_{s},B,\alpha,z)=\sum\frac{(H(H_{0},A_{s},B,\alpha,z)-H_{obs}(z))^2}{\sigma^{2}_{z}}
\end{equation} 
where $H_{obs}(z)$ is the observed Hubble parameter at redshift $z$ and $\sigma_z$ is the error associated with that particular observation  as shown  in Table -\ref{tab3}. 
 Now considering all the observations mentioned above, we consider total chi-square function which  is given by 
\begin{equation}
\chi^{2}_{total}(A_{s},B, \alpha)=\chi^{2}_{f}(A_{s},B,\alpha)+\chi^{2}_{s}(A_{s},B, \alpha)+\chi^2_{H-z}(A_{s},B, \alpha).
\end{equation}
In this case the best fit value is obtained  minimizing  the  chi-square function. Finally we draw  contours at different confidence limits to determine  limiting range of values of the  EoS parameters for the MCG. We now  minimize the  chi-square $\chi^{2}_{f}(A_{s},B, \alpha)$ function with   the growth rate data set. 

\begin{figure}[tbp]
\centering
{\includegraphics[width=230pt,height=190pt]{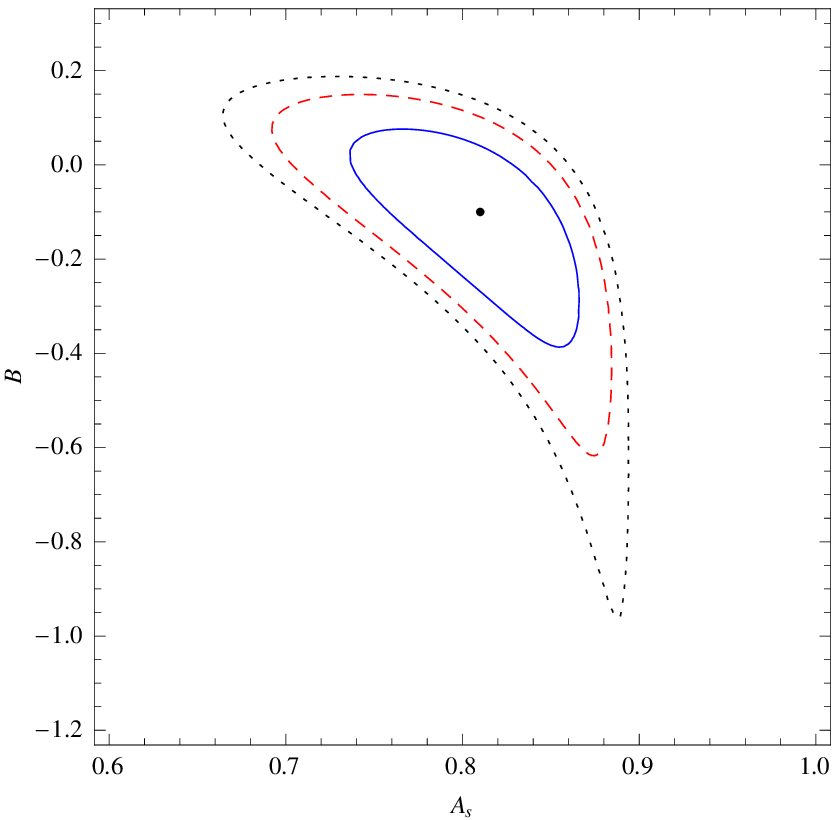}}\\
{\includegraphics[width=230pt,height=190pt]{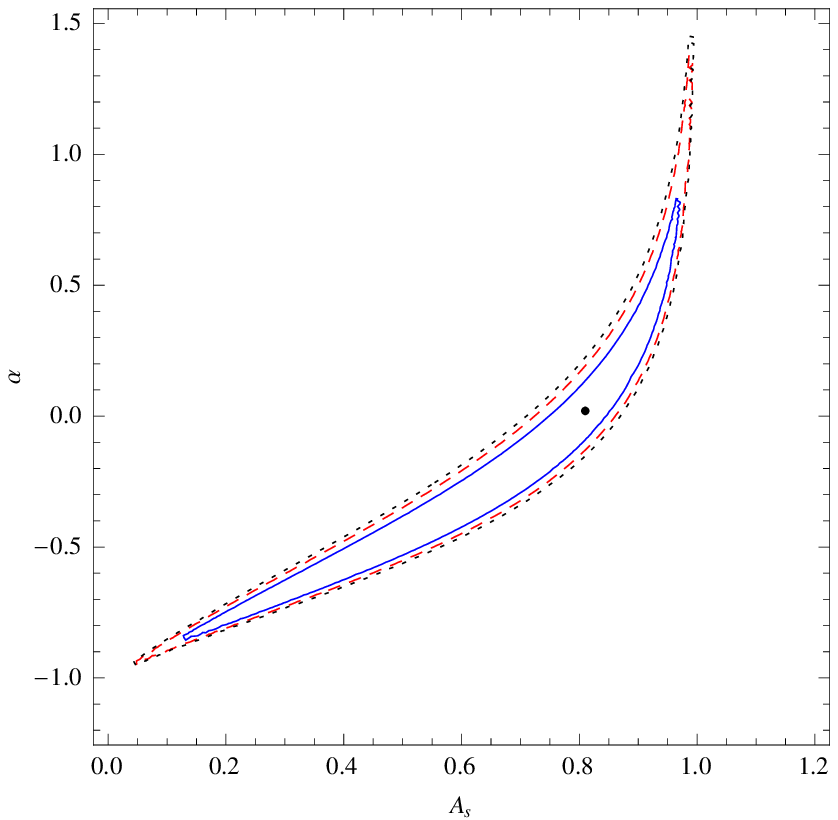}}\\
{\includegraphics[width=230pt,height=190pt]{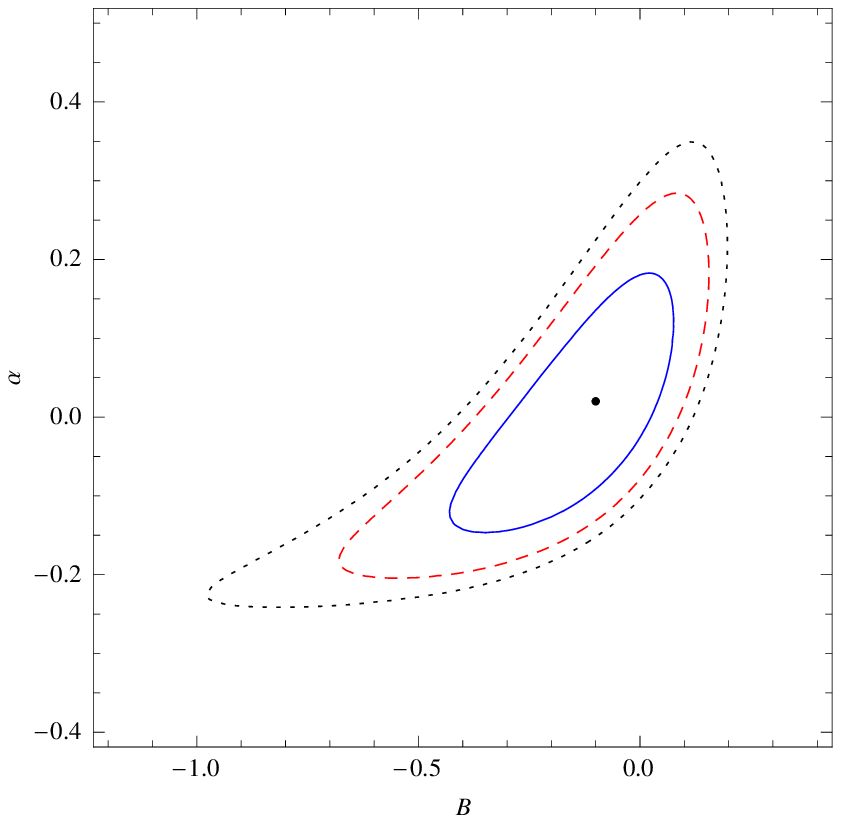}}\\
\caption{\label{contour1} Contours from growth  data  at  $68.3\%$(Solid) $90.0\%$ (Dashed) and $95.4 \%$  (Dotted) confidence limit at best-fit values $A_{s}$=0.81,$B=-0.10$,$\alpha$=0.02.}

\end{figure}

The best-fit values of the  EoS parameters are determined from minimization of chi-squares constituted from  growth function (f),  growth +rms mass fluctuation, growth +rms mass fluctuation+ OHD  separately  which are shown in Table-4. It appears from the analysis that the $A_s$ and $\alpha$ values are least from Growth+rms mass fluctuation+OHD but the parameters are positive. However, the best fit value of $B$  in the later case is a small positive number although a negative value for $B$ is permitted by other two observations.

Using the best fit values for growth data, rowth+ rms mass fluctuation  data and rowth+ rms mass fluctuation  data+OHD from Table-4  we draw contours for (i) $B$ vs. $A_s$ in figs. 1(a), 2(a) and 3(a)  (ii)  $\alpha$ vs. $A_s$ in figs. 1(b), 2(b) and 3(b) and (iii) $\alpha$ vs. $B$ in figs. 1(c), 2(c) and 3(c) respectively. 
Allowed range of values of the EoS parameters  are then determined from the contours. In Table-5  we present allowed range of values of $A_s$ and $B$ (with 95.4 \% confidence limit) for the three cases. It is observed that the range of values for $A_s$ and $B$ are tight. We note that both positive and negative values of $B$ parameters are possible. In Table-6 the range of values of $A_s$ and $\alpha$ are shown. It is evident  from  Growth+ rms mass fluctuation  data+OHD analysis, that the lower bound on $A_s$ is decreased with a tight constraint but both positive and negative values of $\alpha$ are permissible.
 In Table-7 the range of values of $B$ and $\alpha$ are shown. It is evident from  Growth+ rms mass fluctuation  data+OHD analysis that the range of both  $B$  and  $\alpha$ are small. Both positive and negative values of the parameters are admitted.
 
\begin{figure}[tbp]
\centering
{\includegraphics[width=230pt,height=190pt]{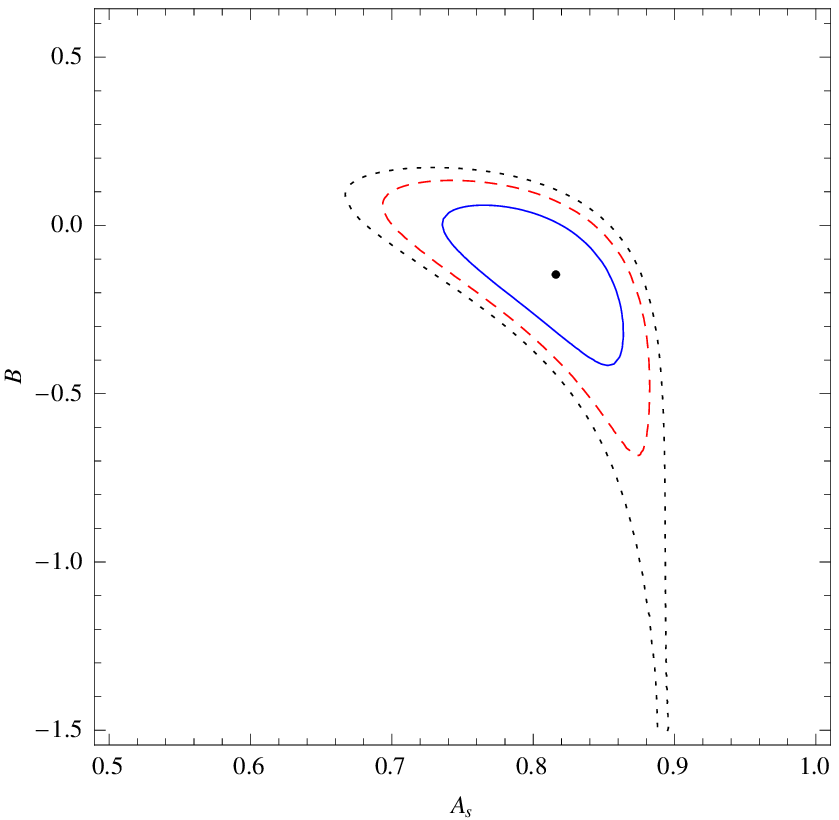}}\\
{\includegraphics[width=230pt,height=190pt]{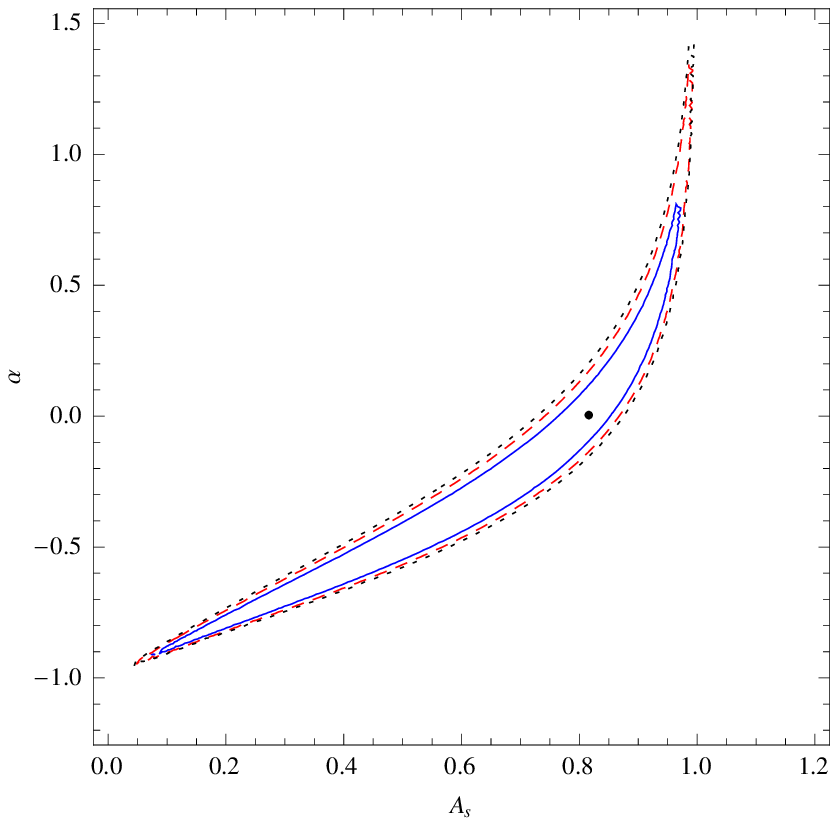}}\\
{\includegraphics[width=230pt,height=190pt]{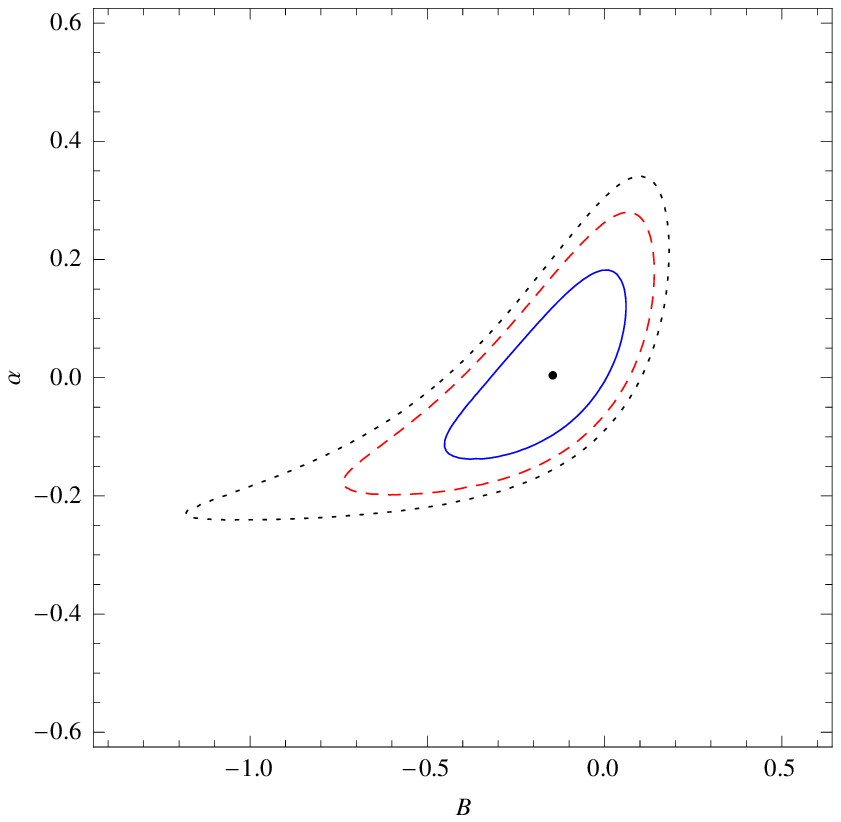}}\\
\caption{\label{contour2} Contours from growth+rms mass fluctuations ($\sigma_{8}$) data  at  $68.3$\%(Solid) $90.0\%$ (Dashed) and $95.4 \%$  (Dotted) confidence limit at best-fit values  :$A_{s}$=0.816,$B$=-0.146,$\alpha$=0.004}
\end{figure}

\begin{figure}[tbp]
\centering
{\includegraphics[width=230pt,height=190pt]{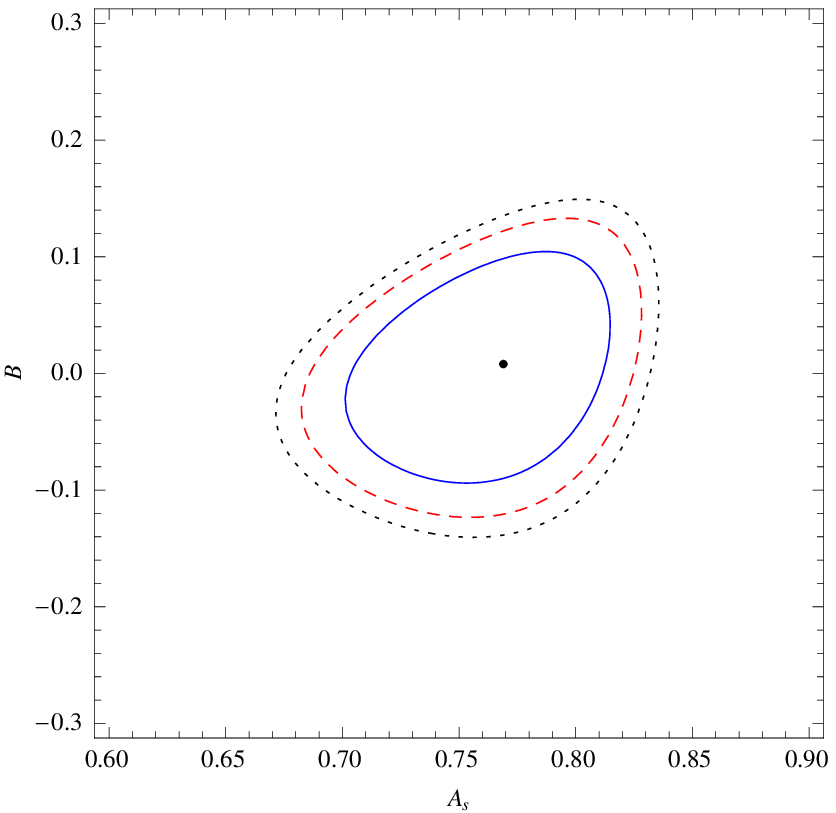}}\\
{\includegraphics[width=230pt,height=190pt]{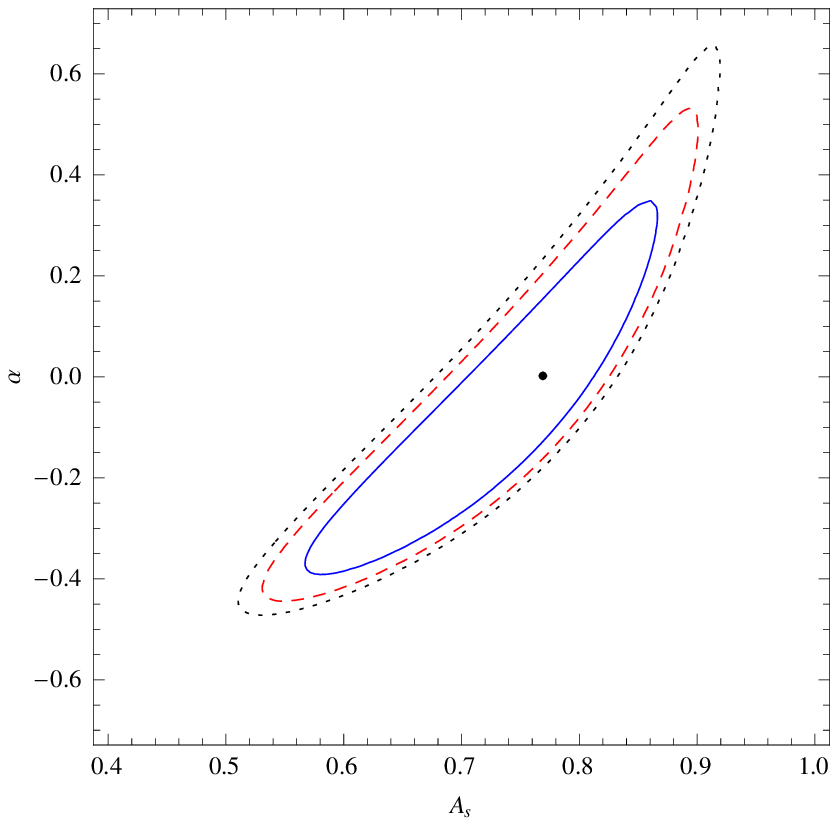}}\\
{\includegraphics[width=230pt,height=190pt]{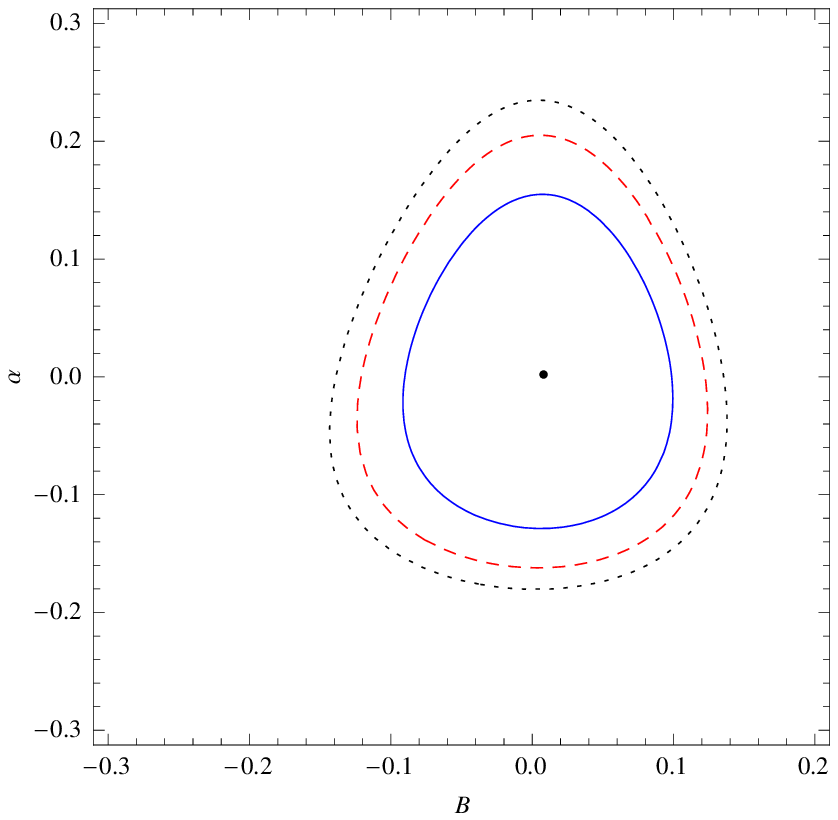}}\\
\caption{\label{contour3} Contours  from growth+rms mass fluctuations ($\sigma_{8}$)+OHD data  at  $68.3\%$(Solid) $90.0\%$ (Dashed) and $95.4 \%$ (Dotted) confidence limit at best-fit values:$A_{s}$=0.769,$B$=0.008,$\alpha$=0.002}
\end{figure}

\begin{figure}
\centering
{\includegraphics[width=230pt,height=190pt]{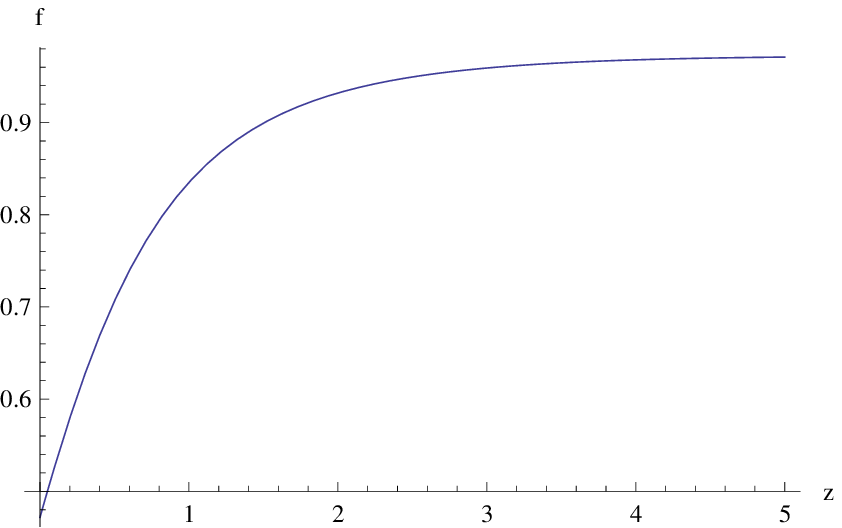}}\\
\caption{\label{growth1} Evolution of growth function f with redshift at best-fit values :$A_{s}$=0.769,$B$=0.008,$\alpha$=0.002.}
\end{figure}

In fig. (4) the growth function $f$  is plotted with redshift $z$ using best fit values of model parameters. It is evident that the growth function $f$ lies in the range 0.472 to 1.0 for redshift between $z=0$ to $z=5$. Initially $f$ remains a constant but it falls sharply  at low redshifts, indicating the fact that major growth of our universe have occurred at the early epoch with  moderate redshift value.

\begin{figure}
\centering
{\includegraphics[width=230pt,height=190pt]{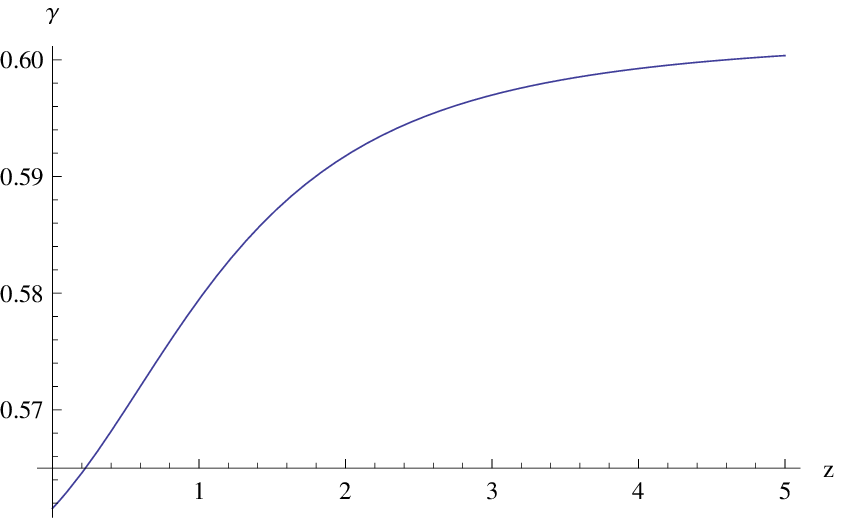}}\\
\caption{\label{growth index } Evolution of growth index $\gamma$ with redshift at best-fit values :$A_{s}$=0.769,$B$=0.008,$\alpha$=0.002.}
\end{figure}

In fig. (5) we plot the variation of the growth index ( $\gamma$ ) with redshift ($z$). It is evident that the  growth index ( $\gamma$) varies between 0.562 to 0.60 for a variation of  redshift between $z=0$ to $z=5$. Thus we observe a  sharp fall in the values  of $\gamma$ at low redshift.

\begin{figure}
\centering
{\includegraphics[width=230pt,height=190pt]{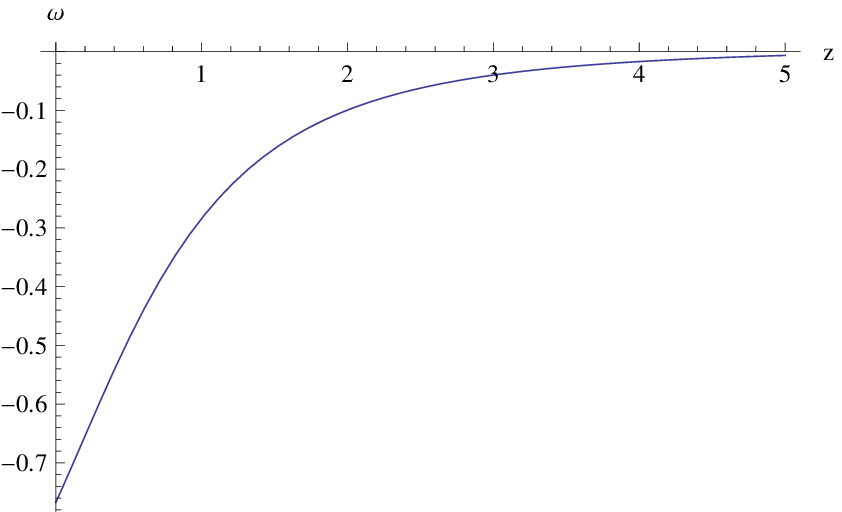}}\\
\caption{\label{state parameter of the EOS} Evolution of the state parameter ($\omega$) at best-fit values:$A_{s}$=0.769,$B$=0.008,$\alpha$=0.002}
\end{figure}

In fig. (6) the variation of the   state parameter ($\omega$) is plotted with $z$. It is found that the  state parameter ($\omega$) varies from  -0.767  at the present epoch  ($z=0$) to $\omega \rightarrow 0$ at intermediate redshift ($z=5$). This result indicates that the universe is now passing  through an accelerating phase which  is dominated by dark energy
whereas in the early universe ($z> 5$) it was dominated by matter permitting a decelerating phase.

\begin{figure}
\centering
{\includegraphics[width=230pt,height=190pt]{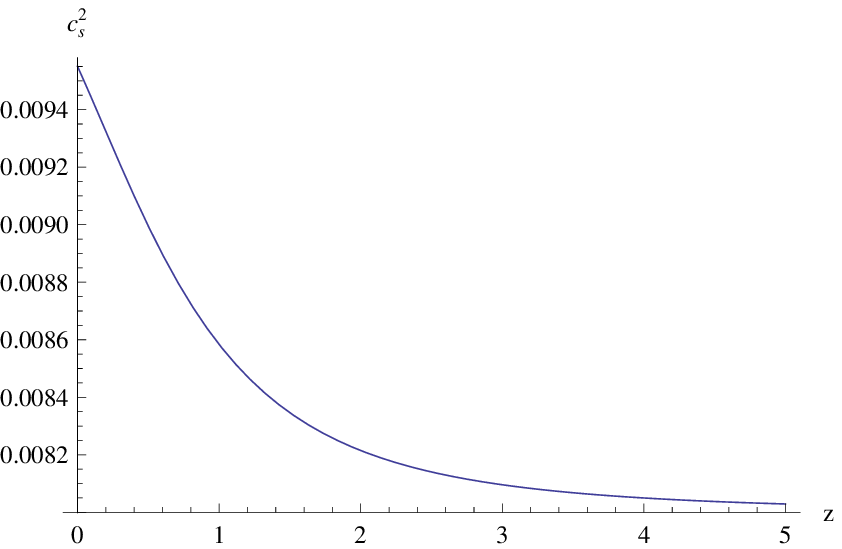}}\\
\caption{\label{sound speed} Sound speed variations with redshift at best-fit values :$A_{s}$=0.769,$B$=0.008,$\alpha$=0.002}
\end{figure}

The variation  of  sound speed $c_{s}^2$ with redshift ($z$)  is plotted in fig. (7). It is noted that $c_s^2$  varies  between 0.0095 to 0.0080 in the above  redshift range admitting causality. A small positive value indicates the occurance of growth in the structures of the universe.

\begin{table}[tbp]
\centering
		\begin{tabular}{|lr|c|c|}
		\hline
		Data     & $A_{s}$ &    $B$ &    $\alpha$\\
		\hline
		$Growth$   & 0.810 &  -0.100 &  0.020\\
		$Growth+\sigma_{8}$ & 0.816 &  -0.146 &  0.004\\
		$Growth+\sigma_{8}+OHD$ & 0.769 &  0.008 &  0.002\\
	
	  \hline	
		\end{tabular}
\caption{\label{tab4} Best-fit values of the EoS parameters}		
\end{table}

\begin{table}
\centering
		\begin{tabular}{|lr|c|c|}
		\hline
		Data &  $CL$  & $A_{s}$ &  $B$ \\
		\hline
		$Growth$  & $95.4\%$   &  $(0.6638,\; 0.8932)$ & $(-0.9758, \;  0.1892)$  \\
		$Growth+\sigma_{8}$  & $95.4\%$ & $(0.6649, \; 0.896)$ & $(-1.5000, \;  0.1765) $\\
		$Growth+\sigma_{8}+OHD$ & $95.4\%$  &  $(0.6711, \; 0.8346)$ & $(-0.1412, \;  0.1502)$\\
	\hline	
		\end{tabular}
\caption{\label{tab5} Range of  values of the EoS parameters}		
\end{table}

\begin{table}
\centering
		\begin{tabular}{|lr|c|c|}
		\hline
		Data &  $CL$  & $A_{s}$ & $\alpha$ \\
		\hline
		$Growth$ & $95.4\%$   &  $(0.0497, \; 0.9935)$ & $(-0.9469, \;  1.460)$  \\
		$Growth+\sigma_{8}$  & $95.4\%$    & $(0.0458, \; 0.9975)$ & $(-0.9469, \; 1.442) $\\
		$Growth+\sigma_{8}+OHD$ & $95.4\%$  &  $(0.5094, \; 0.9204)$ & $(-0.4770, \;  0.6562)$\\
	
	  \hline	
		\end{tabular}
\caption{\label{tab6} Range of  values of the EoS parameters}		
	\end{table}
	
\begin{table}
\centering
		\begin{tabular}{|lr|c|c|}
		\hline
		Data &  $CL$  & $B$ & $\alpha$ \\
		\hline
		$Growth$  & $95.4\%$   &  $(-0.9764, \; 0.1979)$ & $(-0.2439 , \; 0.3525)$  \\
		$Growth+\sigma_{8}$ & $95.4\%$    & $(-1.186, \; 0.2754)$ & $(-0.2436, \; 0.3423) $\\
		$Growth+\sigma_{8}+OHD$ & $95.4\%$  &  $(-0.1449, \; 0.1386)$ & $(-0.1818, \; 0.2360)$\\
		 \hline	
		\end{tabular}
\caption{\label{tab7} Range of  values of the EoS parameters}	
	\end{table}

\begin{table}
\centering
		\begin{tabular}{|lr|c|c|c|c|c|c|}
		\hline
		Model &   $A_{s}$ &  $B$ & $\alpha$ & $f$ & $\gamma$&$\Omega_{m0}$& $\omega_{0}$\\
		\hline
		$MCG$ & $0.769$   &  $0.008$ & $0.002$ & $0.472$& $0.562$& $0.262$& $-0.767$\\
		        
		$GCG$ & $0.708$   &  $0.0$ & $-0.140$ & $0.477$& $0.564$& $0.269$& $-0.708$ \\
		                    
		$\Lambda CDM$ &  $0.761$   &  $0.0$ & $0.0$ & $0.479$& $0.562$& $0.269$& $-0.761$ \\
		                        
	  \hline	
		\end{tabular}
\caption{\label{tab8} Values of the EoS parameters in different model}	
	
\end{table}

\section{Discussion}

We present  cosmological models with MCG as a candidate for dark energy and determine the allowed range of values of the EoS parameters making use of observed data. We also study the growth of perturbation for large scale structure formation in this model using modified Chaplygin gas.  
 The observed data sets are used to study the growth of matter perturbation in MCG model and determined  the range of  growth index $\gamma$ as per \cite{wangstein} in terms of MCG parameters  following the method adopted in \cite{anjan}. The model parameters are constrained  using  the  observational data from redshift distortion of galaxy power spectra and the $rms$  mass fluctuation ($\sigma_{8}$) from Ly-$\alpha$ surveys. It may be pointed out here that the redshift distortion is  interesting and now-a-days leads to an exciting prospect regarding the testing of  different gravity models. The compiled data set  consists 11 data points in growth data set given in  Table-\ref{tab1}  including the four latest Wiggle-Z survey data \cite{blake}  are used for the analysis. There are 17 data points shown in   Table-\ref{tab2}  which are used to study growth rate in addition to  $\sigma_{8}$ data from the power spectrum of Ly-$\alpha$ surveys. We also use Stern data set \cite{stern} corresponding to $ H(z)$  vs.  $z$  data ( Table-\ref{tab3}) for our analysis.

The best-fit values of the parameters $A_{s}$, $B$, $\alpha$ obtained from $\chi^{2}_{tot}(A_{s}$,B,$\alpha)$ are   shown in table(\ref{tab4}).
Using the best fit values we analyse the model and obtained the allowed range of values of the EoS parameters which  are shown in Tables-\ref{tab5},\ref{tab6} and \ref{tab7} respectively.

The best-fit values of the growth parameters for MCG at the present epoch ($z=0$) are  $f$=0.472 ,  $\gamma$=0.562,  $\omega$=-0.767 and $\Omega_{m0}$=0.262 (shown in Table-\ref{tab8}).
It is also noted that the growth function $f$ varies between 0.472 to 1.0  and the growth index  $\gamma$  varies between 0.562 to 0.60 for a variation of redshift from $z=0$ to $z=5$.
 In this case the state parameter $\omega$ lies between -0.767 to 0 with sound speed $c_{s}^2$ that varies between  0.0095 to 0.0080. \\

Thus we note that a satisfactory cosmological model emerges permitting present accelerating universe with MCG in GTR.
The negative values of state parameter ($\omega \leq$ -1/3) signifies the existence of such a phase of the universe. The sound speed obtained in the model is also small which permits structure formation. Thus the MCG model is a good candidate for describing evolution of the universe which  reproduces the  cosmic growth with inhomogeneity in addition to  a late time accelerating phase.   

In   Table-\ref{tab8}  we present  values of the EOS parameters, present growth parameters and density parameter ($\Omega$)  for MCG, GCG, $\Lambda$CDM model. It is evident that the  observational constraints that puts on MCG model parameters are close to $\Lambda$CDM model than GCG model. The MCG model reduces to GCG for $B=0$ and  $\Lambda$CDM model for $B=0$  and $\alpha=0$. MCG model is considered as a good fit model  with recent cosmological observations accommodating recent acceleration followed by a decelerating phase.

\section{Acknowledgements}
PT would like to thank {\it IUCAA Reference Centre} at North Bengal University for extending necessary research facilities to initiate the work. BCP would likt to thank UGC for MRP (2013),

\end{document}